\documentclass[10pt,twocolumn,twoside]{IEEEtran}
\IEEEoverridecommandlockouts

\usepackage{amssymb}
\usepackage{cite}
\usepackage{graphicx}
\usepackage{array}
\usepackage{multirow}
\usepackage{amsmath}
\usepackage{stfloats}
\usepackage{tabularx}
\usepackage{epsfig,epsf}
\usepackage{setspace}
\usepackage{bm}
\usepackage{textcomp}
\usepackage{algorithmic}
\usepackage{algorithm}
\usepackage{caption}
\usepackage{setspace}
\usepackage{url}
\usepackage{verbatim}
\usepackage{color}
\usepackage{xpatch}
\usepackage{hyperref}
\usepackage{subfloat}
\usepackage{subcaption}
\usepackage{xcolor}
\usepackage{colortbl}

\definecolor{shadecolor}{RGB}{0,0,255}
\definecolor{blue}{RGB}{0,0,255}

\begin{document}

\title{Enhanced Ground–Satellite Direct Access via Onboard Rydberg Atomic Quantum Receivers}

\author{Qihao Peng, Tierui Gong,~\IEEEmembership{Member,~IEEE}, 
 Zihang Song,~\IEEEmembership{Member,~IEEE}, 
 Qu Luo,~\IEEEmembership{Member,~IEEE},\\ 
Zihuai Lin,~\IEEEmembership{Senior Member,~IEEE}, 
Pei Xiao,~\IEEEmembership{Senior Member,~IEEE}, 
Chau Yuen,~\IEEEmembership{Fellow,~IEEE}.

		\thanks{Q. Peng, Q. Luo, and P. Xiao are affiliated with 5G and 6G Innovation Centre, Institute for Communication Systems (ICS) of the University of Surrey, Guildford, GU2 7XH, UK. (e-mail: \{q.peng,q.u.luo, p.xiao\}@surrey.ac.uk). (\emph{Corresponding authors: Tierui Gong} and \emph{Zihang Song}). } 
        \thanks{T. Gong and C. Yuen are with the School of Electrical and Electronics Engineering, Nanyang Technological University, Singapore 639798 (e-mail: trgTerry1113@gmail.com, chau.yuen@ntu.edu.sg). 
        } 
        \thanks{Z. Song  is with  Connectivity (CNT) Section, Department of Electronic Systems, Aalborg University, 9220 Aalborg, Denmark (e-mail: zsong@es.aau.dk).} 
        \thanks{Z. Lin is with the School of Electrical and Information Engineering, The University of Sydney, Sydney, NSW 2006, Australia. (e-mail: zihuai.lin@sydney.edu.au).} 
      
        }

\maketitle

\begin{abstract}
    Ground-satellite links for 6G networks face critical challenges, including severe path loss, tight size‑weight‑power limits, and congested spectrum, all of which significantly hinder the performance of traditional radio frequency (RF) front ends. This article introduces the Rydberg Atomic Quantum Receiver (RAQR) for onboard satellite systems, a millimetre‑scale front end that converts radio fields to optical signals through atomic electromagnetically induced transparency. RAQR’s high sensitivity and high frequency selectivity \textcolor{black}{have the potential to} address link‑budget, payload, and interference challenges while fitting within space constraints. \textcolor{black}{Theoretically}, \textcolor{black}{a hybrid atomic–electronic design that is supported by a consistent signal model achieves spectral efficiency exceeding 6 bit/s/Hz, extends coverage by up to 1000 km, and improves sensing accuracy by two orders of magnitude relative to conventional RF receivers. The paper concludes with integration strategies, distributed‑satellite concepts, and open research challenges for bringing RAQR‑enabled satellite payloads into service.} 
\end{abstract}	
	
	

\section{Introduction}

The demand for global coverage in sixth-generation (6G) communication systems, coupled with significant private investment in the space sector, positions satellite communication as a strategic pillar for ubiquitous and direct connectivity \cite{yue2023low}. To this end, Third Generation Partnership Project (3GPP) has integrated satellite links into its non-terrestrial network standard, fostering a vision of seamless terrestrial-satellite network integration. Achieving the full potential of this integrated network relies heavily on robust and efficient ground-to-satellite direct link, which serves as the crucial conduit for large coverage data transmission and wireless sensing. 

High-throughput satellite direct connectivity faces escalating demand. This demand stems from emerging mission profiles like real-time Earth observation, broadband backhaul, and large-scale Internet of Things integration. To meet the performance requirements of these data-intensive applications, contemporary satellite systems increasingly adopt multi-input multi-output (MIMO) architectures. This architectural shift leverages spatial multiplexing \cite{you2020massive} and is evident in current high-throughput satellite implementations. However, as shown in Table \textcolor{red}{I}, the implementation of MIMO for ground-to-satellite direct link faces unique challenges that differ from terrestrial MIMO systems due to inherent constraints as follows.

\emph{Link Budget Constraints:}
Long propagation distances result in significant free-space path loss (FSPL), typically 150-200 dB \cite{heo2023mimo}. Atmospheric effects, including rain, gaseous absorption, and particularly turbulence, severely weaken direct links. Turbulence causes more pronounced beam spreading and distortion compared to downlinks because the beam traverses the densest part of the atmosphere early in its path. Further losses from antenna misalignment and pointing errors collectively narrow the link margin. Therefore, ensuring reliable communication through direct ground-to-satellite links remains a significant challenge.

\emph{Payload Resource and Integration Constraints}:
Satellite payloads operate under severe size, weight, and power (SWaP) limitations \cite{xiao2022antenna}. This directly impedes the deployment of large phased arrays and full digital beamforming architectures for reception. Such systems demand extensive RF circuitry, high power consumption, thermal management, and robust, radiation-hardened components. Integrating scalable MIMO hardware within the strict SWaP envelopes, driven by launch costs and orbital sustainability, remains a significant hurdle for satellite receivers.

\emph{Spectrum Congestion and Interference Vulnerability}:
Satellite frequency bands are heavily shared among civil, commercial, and governmental users (e.g., L, S, C, Ku bands). To increase throughput, aggressive frequency reuse strategies are common. These strategies, in turn, introduce substantial co-channel and adjacent-channel interference from other ground stations or satellite systems \cite{kodheli2020satellite}. Moving to higher frequency bands (e.g., Ka or Q/V bands) offers more bandwidth but brings new design challenges for onboard receivers. These challenges include increased rain fade, stricter hardware linearity requirements, and elevated thermal instability in front-end components. Furthermore, ground-based jamming poses a direct threat to uplink integrity, which requires onboard receivers with robust interference rejection.

Overcoming these intertwined constraints calls for a receive architecture on the satellite that combines very high sensitivity, compact physical dimensions, and minimal power consumption while maintaining interference tolerance. This motivates the investigation of radically different front-end architectures capable of operating under extreme power, size, and spectral isolation constraints, particularly for the reception of ground-originated signals. Quantum-enhanced sensing technologies represent a particularly promising direction. One such approach is the Rydberg atomic quantum receiver (RAQR), which utilizes highly excited Rydberg states of atoms to transduce RF fields from the ground segment into optical signals \cite{gong2024rydberg}. These receivers can directly demodulate RF signals via electromagnetically induced transparency (EIT) without requiring conventional components such as local oscillators or RF mixers. Their unique properties, such as wide frequency tunability, ultra-high field sensitivity to weak signals, inherent optical isolation from thermal noise, and the potential for chip-scale integration, make RAQRs an intriguing candidate for next-generation spaceborne front-ends \cite{gong2024rydberg}.

This paper introduces the concept of Rydberg-based RF sensing to the satellite communications community, focusing on its potential to mitigate the key uplink challenges identified earlier. We first review the physical principles and signal processing mechanisms underlying atomic RF detection. We then evaluate its suitability under satellite-specific constraints such as link budget, SWaP, and spectral congestion for uplink reception. Finally, we propose an architecture for hybrid atomic–electronic satellite receivers and outline future research directions to bridge quantum physics with scalable aerospace communication systems.

\begin{table*}[t]
    \normalsize
    \centering
    \caption{\textcolor{black}{RAQR Capabilities Relevant to Satellite Challenges}}
    \renewcommand{\arraystretch}{1.3}
    \begin{tabular}{|p{4cm}|p{5cm}|p{7cm}|}
        \hline
        \rowcolor{cyan!10} \textbf{Satellite Communication Challenges} & \textbf{RAQR-Based benefits} & \textbf{Practical challenges and potential solutions}\\
        \hline
        \textbf{Link Budget Constraints:} Severe FSPL (150 - 200 dB), constrained power, and limited antenna gain. & 
        \textcolor{black}{\textbf{\textcolor{black}{High Sensitivity}}: Ultra-high dipole moment between Rydberg transitions and limited thermal noise impacts \cite{gong2024rydberg,bussey2022quantum}.}&
        \textcolor{black}{\textbf{\textcolor{black}{Challenges}}: Experimental sensitivity related to atomic density, temperature, detection, and laser drift.} \par
        \textcolor{black}{\textbf{\textcolor{black}{Solutions}}: Frequency-locked lasers, ovenized cell, and thermal management.} \\
        \hline
        \textbf{Payload SWaP Constraints}: Constrained volume and mass budgets, limited solar panel area, and power generation. &
       \textcolor{black}{\textbf{Miniaturized Architecture}: Flexible extension for multi-functional receivers using a single vapor cell, wavelength-independent sensing volume, and no traditional RF frontend  \cite{gong2024rydberg}.} &
       \textcolor{black}{\textbf{Challenges}: Stabilized laser provisioning, micro vapor-cell packaging and alignment, and frequency calibration.} \par
        \textcolor{black}{\textbf{\textcolor{black}{Solutions}}: Centralized stabilized master laser, module multiplexing, and micro-integrated waveguide .} \\
        \hline
        \textbf{Spectrum Congestion and Interference}: Aggressive frequency reuse, dense orbital constellations, and adjacent-channel leakage. & 
        \textcolor{black}{\textbf{Tunable Spectral Selectivity}: Laser-controlled energy level coupling, Rydberg state-dependent dipole transitions, and sub-kHz spectral resolution \cite{anderson2020rydberg}.} &
       \textcolor{black}{\textbf{Challenges}: Limited instantaneous bandwidth and discrete frequency coverage.} \par
        \textcolor{black}{\textbf{\textcolor{black}{Solutions}}: Multi-band or multi-carrier transmission, multi-species/multi-series state, and atomic heterodyne mode.} \\
        \hline
        \textbf{Hardware Instability}: Thermal drift under vacuum conditions and oscillator drift. &
        \textcolor{black}{\textbf{Quantum Characteristics}: Quantum-coherent field transduction and atomic energy levels with thermal insensitivity} &
         \textcolor{black}{\textbf{\textcolor{black}{Challenges}}: Temperature, frequency drift, blackbody radiation-induced dephasing, and power broadening.}\par
        \textcolor{black}{\textbf{\textcolor{black}{Solutions}}: Multi-zone ovenized cell, automatic drift compensation, and periodic recalibration.} \\
        \hline
    \end{tabular}
    \label{tab:raqrs_satellite_challenges}
    \vspace{-0.6em}
\end{table*}

\section{RAQRs for ground-satellite link}

\subsection{Fundamentals of Rydberg Atomic Quantum Receivers}

In this subsection, we present the fundamentals of RAQRs, including their quantum sensing principles, representative implementations, typical photodetection schemes, mathematical theories, and natural benefits for satellite applications.

\subsubsection{Quantum Sensing Principles}
\begin{figure*}
    \centering
    \includegraphics[width=0.98\linewidth]{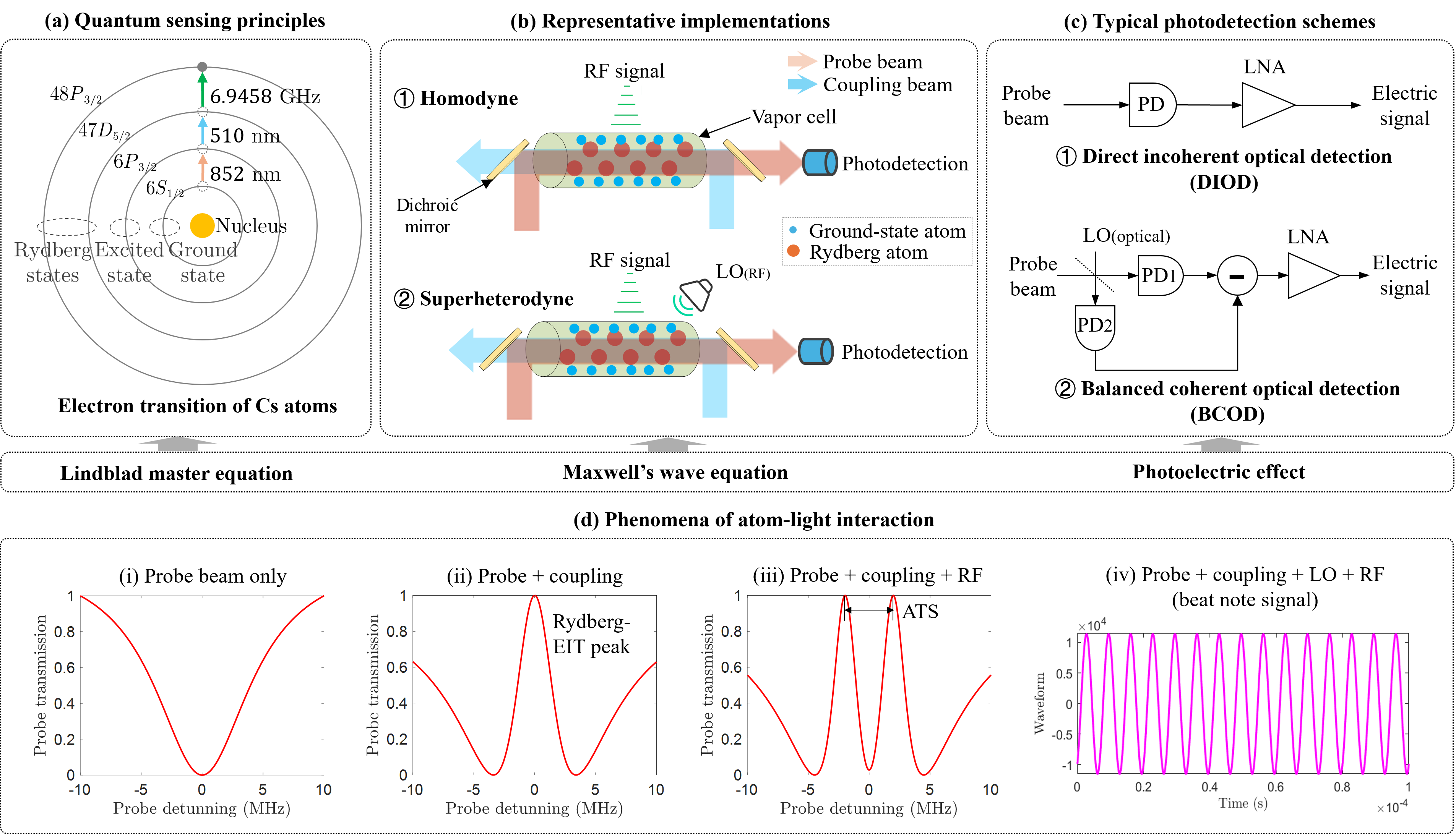}
    \caption{Fundamentals of RAQRs}
    \label{fig:Fundamentals}
    \vspace{-1.0em}
\end{figure*}

RAQRs exploit an ensemble of Rydberg atoms for the purpose of ultra-sensitive RF field detection by harnessing their extremely high dipole moment. Rydberg atoms are created from hydrogen-like alkali (Cs or Rb) atoms, where the outmost electron of an alkali atom is excited from the ground state to a Rydberg state. This excitation is realized by exploiting two or three electron transitions when these atoms absorb external electromagnetic waves having a specific frequency coupled with the corresponding energy levels. For example, as shown in Fig. \ref{fig:Fundamentals}(a), the first electron transition of an Cs atom happens between the ground state 6S\textsubscript{\scalebox{0.8}{1/2}} and an excited state 6P\textsubscript{\scalebox{0.8}{3/2}} by absorbing a $852$ nm laser beam, followed by a second electron transition from 6P\textsubscript{\scalebox{0.8}{3/2}} to the Rydberg state 47D\textsubscript{\scalebox{0.8}{5/2}} by absorbing a $510$ nm laser beam. These processes constitute the quantum state preparation of RAQRs for further RF detection. 

When a desired RF field (e.g., having a carrier frequency of 6.9458 GHz) is impinging to the Rydberg atoms, a further electron transition occurs between two Rydberg states (e.g., from 47D\textsubscript{\scalebox{0.8}{5/2}} to 48P\textsubscript{\scalebox{0.8}{3/2}}). This process constitutes the quantum state evolution of RAQRs, where a prepared Rydberg state is time-evolved along with the desired RF field. 
The evolved quantum state is finally measured to offer an access to recover the information of the desired RF field. This quantum state measurement is typically realized by relying on the measurement of an optical signal during an atom-light interaction process, which is detailed in the following section.

\subsubsection{Representative Implementations} 
\textcolor{black}{Fig. \ref{fig:Fundamentals}(b) illustrates the above-mentioned quantum sensing processes of the two-photon excitation scheme.} An ensemble of Rydberg atoms in the form of the atomic vapor is filled up with a glass cell. A pair of laser beams, termed as `probe' and `coupling', counter-propagate through the vapor cell and form a spatially overlapped detection region. Under the joint excitation of the probe and coupling beams, the atoms of this region are excited to the Rydberg state, realizing the quantum state preparation. During this atom-light interaction process, a special but useful physical phenomenon, known as the Rydberg EIT, is generated, which renders a transparent window of the atomic vapor over a narrow range of frequencies of the absorption spectrum. Specifically, by overlapping the probe and coupling beams, the probe beam that should have been absorbed by the atomic vapor becomes transmissible as if the atomic vapor is transparent, as depicted in Fig. \ref{fig:Fundamentals}(d)(ii). 

\textit{Homodyne Scheme}:
The Rydberg-EIT also offers an optical readout manner for measuring the evolved quantum state driven by the desired RF field, where the EIT peak is split into two peaks when the RF field is detected. This phenomenon is known as the Autler–Townes splitting (ATS), where the amplitude of the desired RF field is associated with the splitting distance of the two peaks, as seen in Fig. \ref{fig:Fundamentals}(d)(iii). By exploiting the ATS for the RF signal recovery, this homodyne scheme is only capable of recovering the amplitude of the desired RF field. 

\textit{Superheterodyne Scheme}:
To further support phase recovery of the desired RF field, the superheterodyne scheme was systemically verified in \cite{jing2020atomic}. Compared to the homodyne scheme, an extra strong local RF field, namely the local oscillator (LO), is imposed to Rydberg atoms together with the desired RF field. The superimposed RF field forms an RF beat note, where its amplitude and phase are further embedded into the amplitude of the probe beam, as indicated in Fig. \ref{fig:Fundamentals}(d)(iv), allowing a reverse recovery of both the amplitude and phase of the desired RF field. The superheterodyne scheme exhibits a higher sensitivity and can support modulated signal transmissions, yielding a wider range of applications. 

Indeed, there are also other promising schemes for realizing different functionalities, such as polarization detection, multiband and continuous-band detection, which are built upon the former two typical schemes. Their details can be seen in \cite{gong2025rydberg}, which are omitted in this article.

\subsubsection{Typical Photodetection Schemes} 
RAQRs realize an RF-to-optical transformation, where the information of the desired RF field is transformed into a corresponding optical beam. A subsequent photodetection allows us to record useful optical information that contains information of the desired RF field.  Two photodetection schemes are typically employed, namely the direct incoherent optical detection (DIOD) and balanced coherent optical detection (BCOD), as protroyed in Fig. \ref{fig:Fundamentals}(c). Specifically, the former scheme directly detects the desired optical beam and outputs a corresponding electrical signal by exploiting a photodetector. By contrast, the latter scheme employs an extra strong local optical source together with the desired optical beam to form two distinctive mixing optical signals. They are then detected by two photodetectors, respectively, and are subsequently combined to form a final electrical output. \textcolor{black}{As unveiled in \cite{gong2024rydberg}, the BCOD scheme can approach the photon shot limit by suppressing the proportion of the thermal noise generated by electronic components at the remaining stages after the photodetector, yielding a better performance compared to the DIOD scheme. }

\subsubsection{Mathematical Theories} 
\textcolor{black}{RAQRs lie at the intersection of quantum sensing, atom–light interaction, photodetection, and classical RF/baseband signal processing, requiring a unified modelling framework that bridges microscopic quantum physics with macroscopic signal observables. The modelling can be understood in a stepwise manner. Starting from the Beer–Lambert law, one can describe how the probe laser experiences absorption and phase shift when propagating through an atomic vapor. These absorption and dispersion effects are governed by the medium’s electric susceptibility, which in turn is derived from the Maxwell–Bloch equations linking Maxwell’s electromagnetic wave equation with the atomic density matrix. The atomic density matrix itself evolves according to the Lindblad master equation, which captures the coherent and dissipative dynamics among quantum states. The resulting optical response modulates the probe beam intensity, which is then converted into an electrical signal via the photoelectric effect in the photodetector. Integrating these layers—from quantum coherence to optical response and finally to baseband signal—enables the construction of an equivalent end-to-end signal model for the RAQR transceiver \cite{gong2024rydberg}, providing a clear pathway to analyze system performance and apply advanced signal processing algorithms. }

\begin{figure*}[ht]
        \centering
    \includegraphics[width=0.94\linewidth]{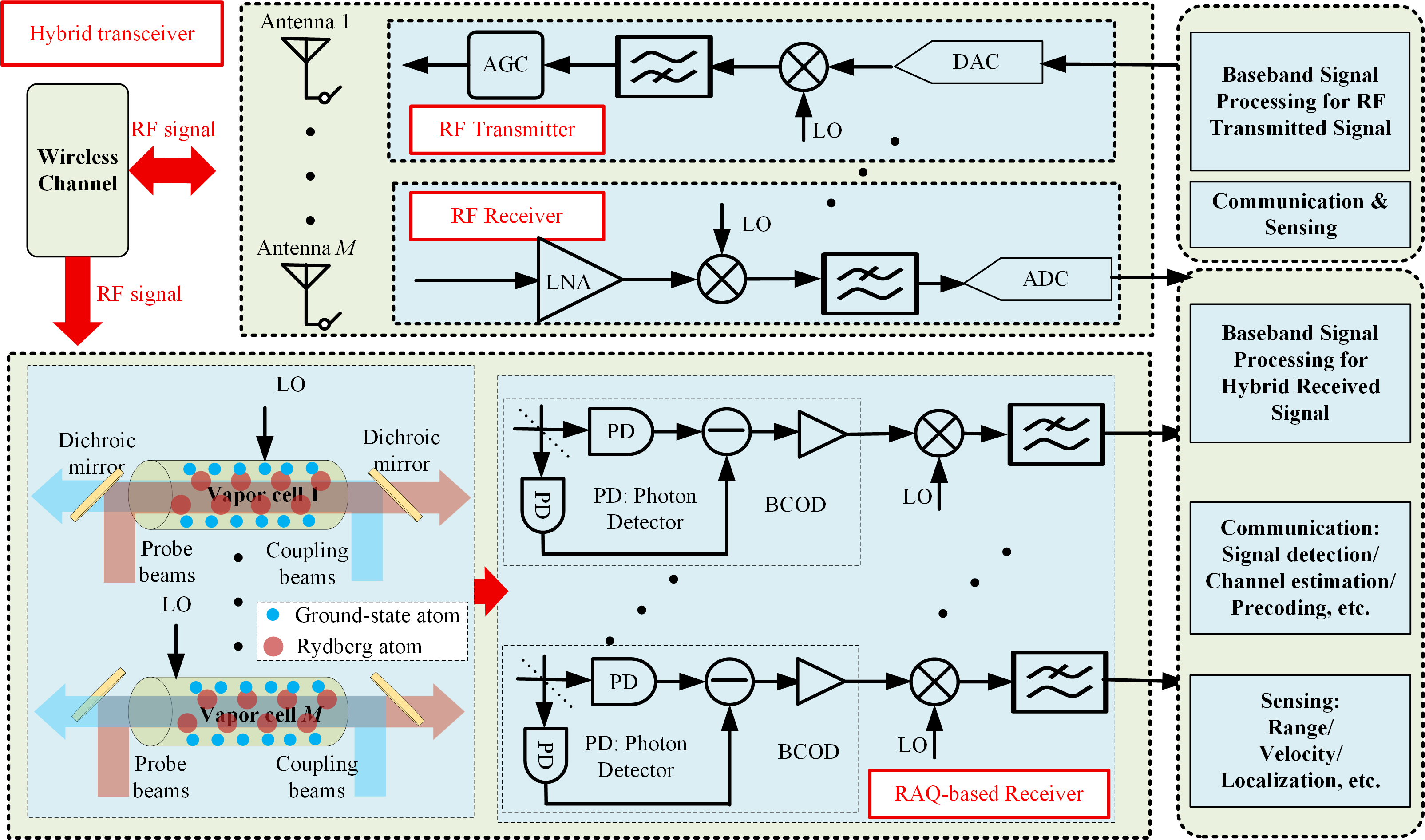}
    \caption{Hybrid Architecture of RF and RQA transceiver}
    \label{hybrid}
    \vspace{-1.0em}
\end{figure*}

\subsection{Natural Benefits of RAQRs for Satellite Applications}
\textcolor{black}{Although integrating the RAQR into a satellite platform remains practically challenging, this approach offers compelling advantages for addressing key satellite communication challenges, primarily through four intrinsic compatibilities, as summarized in Table I.}


 \textcolor{black}{\textbf{High sensitivity}:  Due to practical limitations of the experimental setup, RAQRs did not demonstrate comprehensive superiority over conventional receivers. As revealed in \cite{schlossberger2024rydberg}, the sensitivity highly depends on the experimental configuration, frequency, the detection schemes, and the suppression of practical impairments. In particular, resonant operation enhances the field sensitivity from 23 to 1.59 \(\mathrm{\mu V/cm/\sqrt{Hz}}\) at 10.22 GHz and room temperature \cite{sandidge2024resonant}. Recently, the \(\mathrm{nV/cm/\sqrt{Hz}}\) field-equivalent sensitivities at 8.57 GHz can be achieved by using RF heterodyne detection \cite{wuenhancing}.  Promisingly, the recent theoretical analyses in \cite{gong2024rydberg,bussey2022quantum} and experimental study  \cite{tu2024approaching} have exhibited strong evidence for supporting the superiority of RAQRs.} This exceptional sensitivity implies that RAQR-assisted satellites are capable of directly receiving extremely weak signals from ground users without large antenna apertures or high-gain amplification. 

\textcolor{black}{\textbf{Miniaturized Architecture}: Unlike conventional antennas, the effective size of RAQRs is independent of the signal wavelength, as RAQRs receive RF signals through Rydberg atoms excited by laser beams within a vapor cell \cite{gong2024rydberg}. Existing research has shown that it is possible to replace an RF antenna with a Rydberg atomic sensor. Although components such as laser arrays and frequency-stabilization hardware add mass and volume, these SWaP costs can be reduced through chip-scale photonic integration. Additionally, laser and signal-processing modules can be multiplexed with the spacecraft’s laser communications terminal and can share its stabilized optical platform, power, and thermal management services, thereby mitigating SWaP constraints. As a result, \textcolor{black}{RAQRs have the potential to provide a compact alternative, although meeting SWaP targets is still challenging, especially for the laser subsystem.}}

\textcolor{black}{\textbf{Tunable Spectral Selectivity}: RAQRs offer a highly advantageous blend of uniquely tunable bands and sharp spectral selectivity \cite{anderson2020rydberg}, making them exceptionally suitable for satellite communication applications. By adjusting laser frequencies and engaging specific Rydberg energy levels, these receivers achieve quasi-continuous coverage spanning MHz to THz frequencies with instantaneous sub-kilohertz spectral resolution. This sharp selectivity enables precise differentiation of adjacent-band signals, reducing adjacent-band interference and enhancing spectrum efficiency accordingly. }

 \textcolor{black}{\textbf{Vacuum Benefits}: Even though day-night cycling leads to an unstable temperature for Rydberg atoms, the satellite vacuum environment facilitates stable RAQR operation \cite{kim2016approaching}. Particularly, the natural vacuum space enables easier thermal control, reduces pressure and acoustic perturbations, and lowers gas-collision rates in the vapor cell, which enhances the optical response. Moreover, with appropriate thermal shielding, the interference posed by unexpected radiation can be mitigated, leading to higher sensitivity. In other words, the natural vacuum environment \textcolor{black}{provides a favorable environment that can reduce some perturbations, though practical challenges such as thermal cycling, radiation, and laser stability remain significant}.}
 

\textcolor{black}{Although integration still faces outstanding engineering challenges, including hardware implementation, thermal management, and laser stabilization, \textcolor{black}{these characteristics have the potential to support narrow linewidths and high sensitivity, though achieving this in a space-qualified system remains an open engineering challenge}.}

\section{RAQR-assisted Satellite System Architectures}
In this section, the RAQRs are integrated into classical RF systems by designing the hybrid transceiver and presenting the hybrid processing based on narrow and wide bands, and the potential of RAQR-enabled satellites in both centralized and distributed scenarios is revealed.
\subsection{Hybrid Transceiver Architecture}
\textcolor{black}{Fig. \ref{hybrid} shows the elaborately designed hybrid architecture for the RAQRs and RF transceiver.} Particularly, the devised transceiver can be divided into two parts, including a conventional RF transmitter and a hybrid receiver. A conventional RF transmitter typically consists of a baseband processing unit, a modulator, and a power amplifier. A hybrid receiver consists of two main components: an RF antenna receiver and a RAQR. The RF antenna receiver comprises a low-noise amplifier, a demodulator, and a signal processor, while the RAQR consists of a Rydberg atomic quantum sensor, a photodetection unit, a down-conversion unit, and a signal processor. 

\begin{figure*}[ht]
        \centering
    \includegraphics[width=0.98\linewidth]{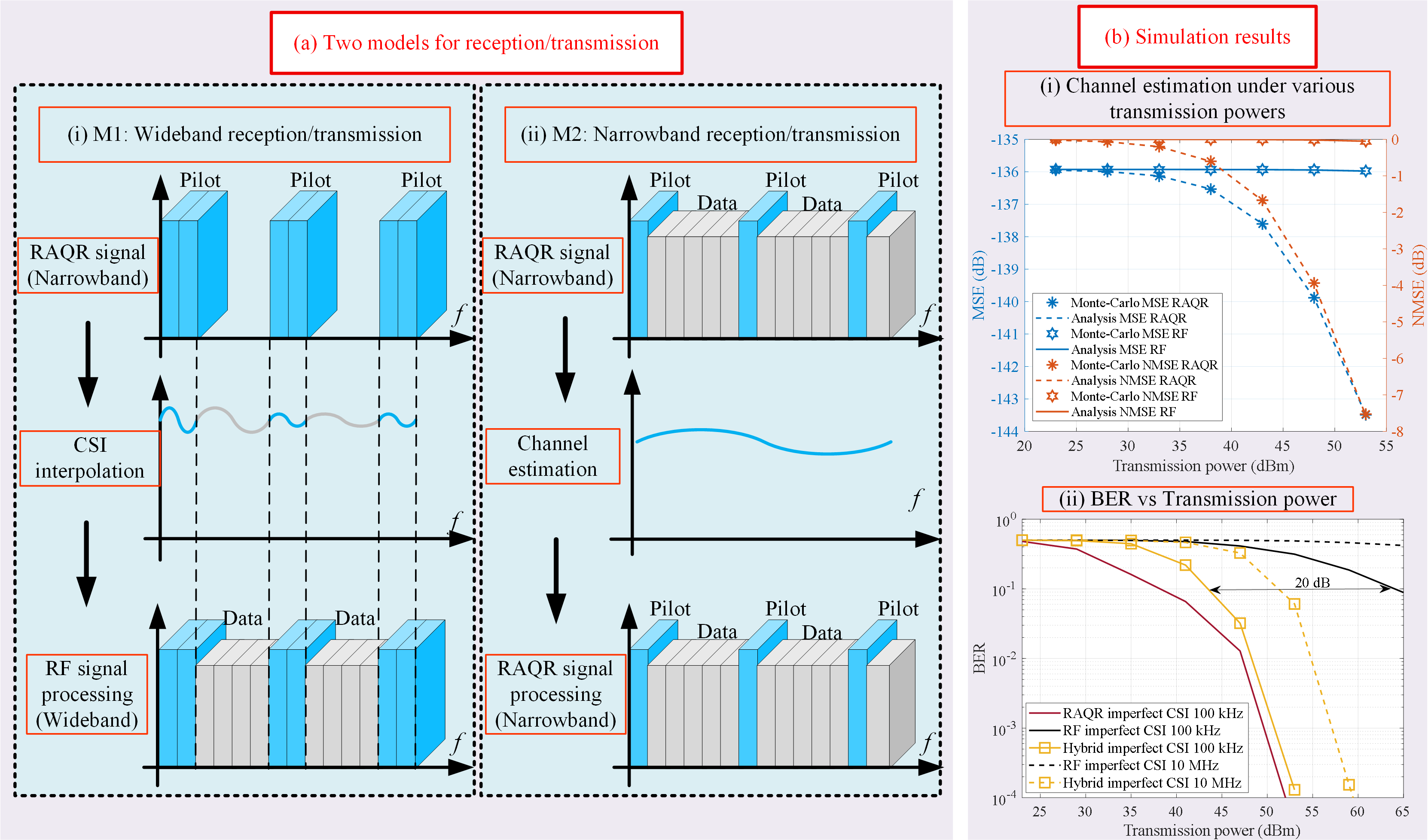}
    \caption{Hybrid signal processing for RAQR-based satellites.}
    \label{workmode}
    \vspace{-1.0em}
\end{figure*}

The equivalent baseband signal is further detailed in \cite{gong2025rydberg}. \textcolor{black}{Based on the received signals from RAQR and the RF receiver simultaneously,  Fig. \ref{workmode}(a)(i) and Fig. \ref{workmode}(a)(ii) illustrate wideband reception and narrowband reception, respectively.}

\textbf{Wideband Reception/Transmission}: Although RAQRs are inherently limited to narrowband reception, they can provide highly accurate channel state information (CSI) within selected frequency bands. By applying interpolation techniques, the partial CSI can be extended across the entire signal bandwidth, thereby improving the performance of conventional RF transceivers in both detection and transmission via hybrid processing. \textcolor{black}{Using the quantum parameters of Table \textcolor{red}{I} in \cite{gong2024rydberg} for the four-level electron transition of  6S\textsubscript{\scalebox{0.8}{1/2}}\textrightarrow 6P\textsubscript{\scalebox{0.8}{3/2}}\textrightarrow 47D\textsubscript{\scalebox{0.8}{5/2}} \textrightarrow 48P\textsubscript{\scalebox{0.8}{3/2}}, and assuming that the noise comes from three parts, including thermal noise, quantum projection noise, and photon shot noise, we evaluate the channel estimation performance by using the minimum mean square error method and then investigate the bit error rate (BER) performance under QPSK modulation, as illustrated in Fig.~\ref{workmode}(b).} Owing to the intrinsic high sensitivity of Rydberg atoms, the accuracy of channel estimation is significantly improved, \textcolor{black}{leading to over 20 dB projected enhancement under the idealized assumptions}. \textcolor{black}{From a theoretical perspective, this result indicates that RAQRs have the potential to address the direct ground-to-satellite connectivity challenge.}

\textbf{Narrowband Reception/Transmission}: Since RAQRs are capable of receiving narrowband signals, the corresponding CSI can be directly applied for signal detection and transmission. Based on the parameters described above, the BER performance is illustrated in Fig.~\ref{workmode}(b), demonstrating an improvement of more than 20 dB over conventional RF receivers and validating the effectiveness of the proposed transceiver for direct satellite-to-ground communication.


\subsection{Scalable RAQR-Based Satellite Networks}
\begin{figure*}[ht]
        \centering
    \includegraphics[width=0.98\linewidth]{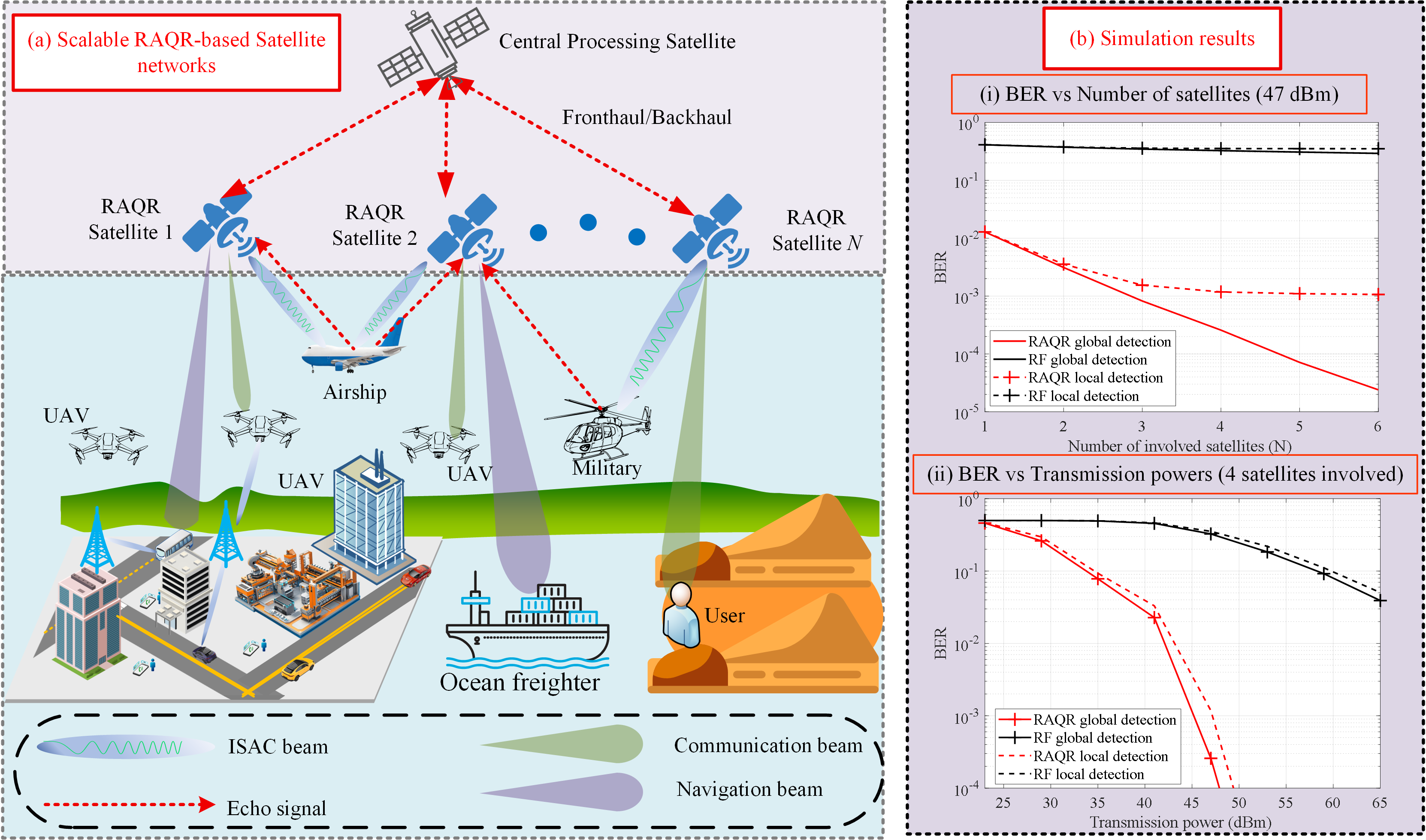}
    \caption{Scalable RAQR-based satellite networks.}
    \label{scalable}
    \vspace{-1.0em}
\end{figure*}

\textcolor{black}{To enhance the spatial diversity and improve the coverage, we further investigate the scalable RAQR-based satellite networks. As shown in Fig.~\ref{scalable}(a), multiple satellites and ground terminals are equipped with RAQRs operating collaboratively within a distributed architecture.} Specifically, each satellite can perform local detection or transmit the received signal to the central processing satellite (head of the dynamic cluster) for global detection.

\textbf{Local Detection}: In local processing, each satellite independently performs signal detection using linear schemes, such as maximum ratio combining or zero-forcing, requiring only local CSI. The decoded information is then forwarded to a selected central satellite via fronthaul. This approach enhances scalability with the number of RAQR-equipped satellites and improves resilience to satellite failures or link disruptions. Additionally, by avoiding the transmission of raw signals or aggregated CSI, local processing substantially reduces fronthaul bandwidth demands.

\textbf{Global Detection}: In centralized global detection, spatially distributed RAQR-enabled satellites independently collect RF signals and perform local channel estimation, subsequently transmitting both raw baseband samples and CSI to a central satellite via a fronthaul. After that, the central satellite performs global signal detection via aggregated CSI and the received signal. This centralized strategy enables coherent combining across distributed satellites, achieving macro-diversity gains that substantially improve interference suppression, spectral efficiency, and detection reliability compared to local processing. Notably, centralized processing can impose higher fronthaul demands compared to local detection.

\textcolor{black}{Using the above-mentioned parameters, Fig. \ref{scalable}(b) provides a theoretical projection of the BER performance for the two detection schemes.} As expected, global detection achieves near-optimal performance by effectively suppressing inter-satellite interference. By contrast, as the number of satellites increases, the BER of local detection tends to plateau. This performance saturation arises from the limited capability of localized processing to mitigate interference originating from distant satellites, thereby leading to a performance bottleneck. Notably, compared to conventional RF-enabled satellite systems, RAQR-based architectures offer enhanced direct ground-to-satellite access and represent a promising solution for next-generation satellite communication networks.

\section{Performance Evaluations of Direct Access}
\textcolor{black}{Next, we present a theoretical projection of integrating RAQRs into classical satellite networks to explore their potential benefits, including possible enhancements in achievable rate, coverage area, and sensing accuracy.}

A four-level electron transition of  6S\textsubscript{\scalebox{0.8}{1/2}}\textrightarrow 6P\textsubscript{\scalebox{0.8}{3/2}}\textrightarrow 47D\textsubscript{\scalebox{0.8}{5/2}} \textrightarrow 48P\textsubscript{\scalebox{0.8}{3/2}} is considered in our simulations, corresponding to the detection of RF signals having a carrier frequency of $f_c = 6.9458$ GHz. The parameters of electron transitions, laser beams, RF signals, and electronic components follow the same values as those listed in TABLE \textcolor{red}{I} of \cite{gong2024rydberg}. \textcolor{black}{These configurations are consistent with the physics experiments and 3GPP specification. Based on these parameter settings, by assuming that there are 100 Rydberg sensors equipped on the satellite, we investigate the benefits of the RAQR-based satellite over a conventional one. Furthermore,} we assume a line-of-sight propagation for the ground-to-satellite transmission. The FSPL is given by $20 \log_{10} \left( c / (4 \pi) \right) + 20 \log_{10} \left( 1/d \right) + 20 \log_{10} \left( 1/f_c \right)$ in dB, where $c$ represents the speed of light and $d$ is the distance. 

\begin{figure*}[t!]
		\centering
		\subfloat[]{
			\includegraphics[width=0.32\textwidth]{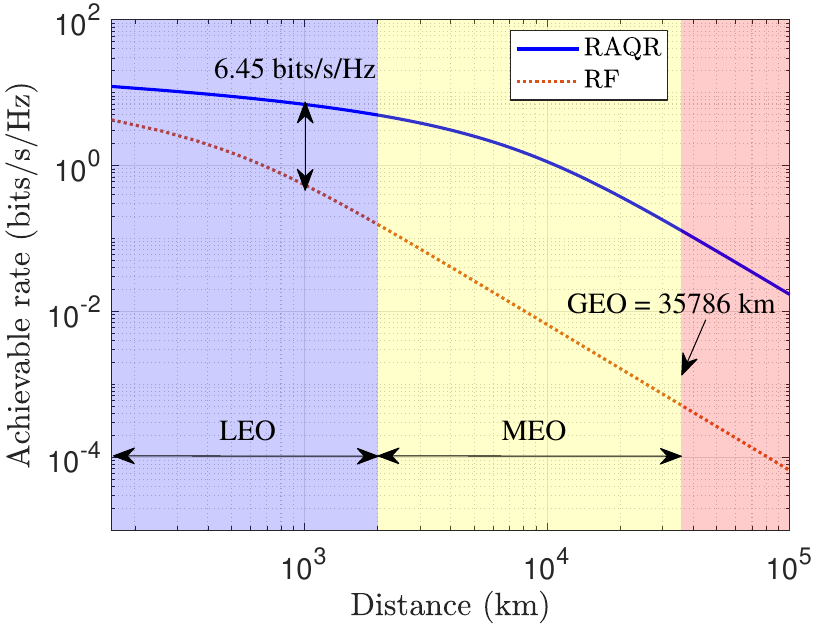}} 
		\subfloat[]{
			\includegraphics[width=0.32\textwidth]{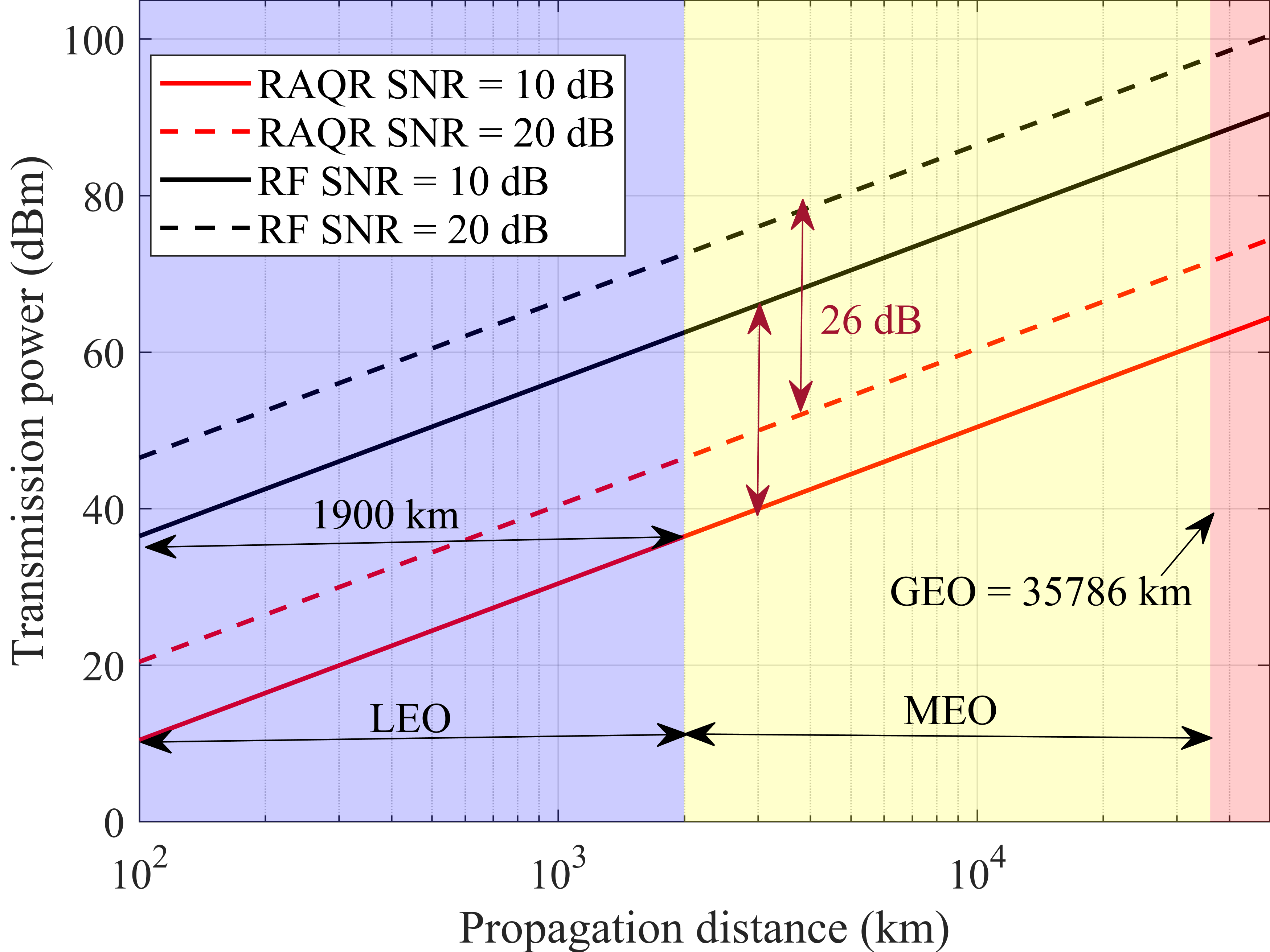}} 
        \subfloat[]{
			\includegraphics[width=0.32\textwidth]{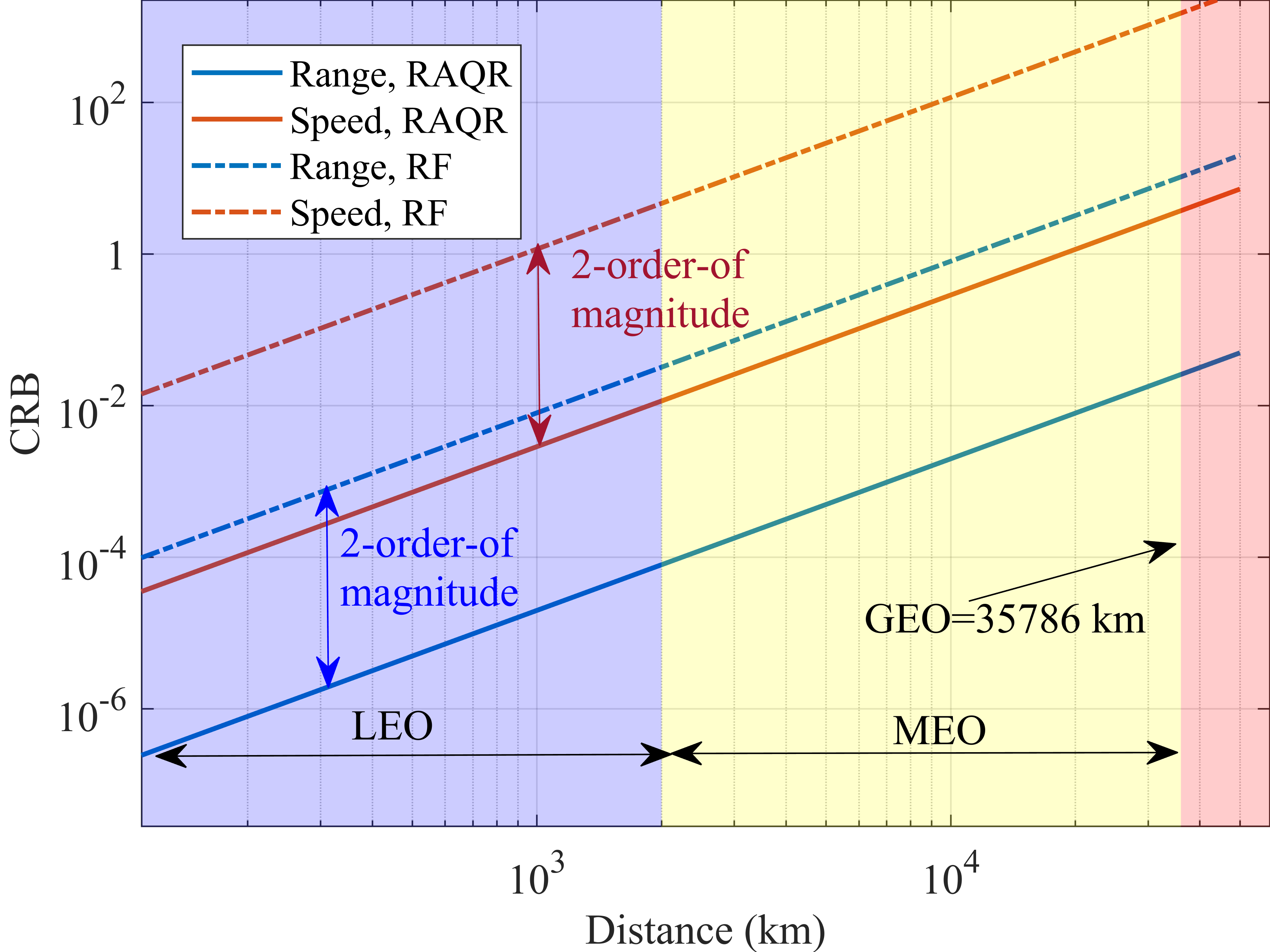}} 
		\caption{\textcolor{black}{Performance of RAQR-enabled satellite systems: (a) achievable rate, (b) coverage area, and (c) sensing accuracy.}}
		\vspace{-1.0em}
		\label{fig:Performance}
\end{figure*}

\subsection{Achievable Rate}
We first verify the achievable rate enhancement of RAQR-enabled satellite systems, as seen in Fig. \ref{fig:Performance}(a). In our simulations, the transmission distance $d$ spans from $160$ km to $2000$ km and from $2000$ km to $35786$ km, covering the range of different types of satellites, such as the low Earth orbit (LEO), medium Earth orbit (MEO) and geostationary equatorial orbit (GEO). As observed from Fig. \ref{fig:Performance}(a), both the RAQR and classical RF receiver experience a degradation of the achievable rate as the distance becomes larger. However, the RAQR exhibits a significant rate enhancement compared to the classical counterpart across the region of the LEO, MEO, and GEO satellites. In particular, the achievable rate can be increased on the order of $6.45$ bits/s/Hz by RAQRs when the transmission distance is $1000$ km. 

\subsection{Coverage Area}

\textcolor{black}{As illustrated in Fig. \ref{fig:Performance}(b), owing to the ultra-high sensitivity, RAQRs enable satellite-based reception of ground-transmitted signals over distances as long as $2000$ km. By comparison, conventional RF antennas operating under the same signal strength constraints (10 dB) are limited to ranges of around $100$ km. Furthermore, for achieving the same received SNR, the energy efficiency relying on RAQR is higher than that based on RF receiver. This substantial extension (around 1900 km) and high efficiency in communication demonstrates the potential of RAQRs to significantly enhance the reliability of the direct satellite-to-ground link, providing a new paradigm for future satellite communication systems.} 

\subsection{Sensing Performance}

 Fig. \ref{fig:Performance}(c)  illustrates the Cramér–Rao bound (CRB) for both range and speed estimation in RAQR-based and RF-based sensing systems, assuming a radar cross section of $15$ dB and $64$ sensing symbols. As observed, the CRB decreases exponentially with increasing sensing distance. Notably, both the range and speed CRBs of the RAQR-based sensing scheme achieve a two-order-of-magnitude improvement over those of the conventional RF-based sensing. These CRB results highlight the promising potential of RAQR technology in satellite sensing applications, where its ultra-high sensitivity enables highly accurate parameter estimation under the challenging conditions of long-range detection.

\section{Directions for Future Research}

\subsection{RAQR Signal Processing and Integration}

Despite its quantum-native advantages, practical deployment of RAQR technology in satellite receivers calls for rethinking the signal processing chain. Future research should investigate hybrid receiver architectures that combine the RAQR’s ultra-sensitive field detection with conventional RF components to expand operational bandwidth and ensure compatibility with existing modulation formats. Methods for dynamic frequency agility, signal fusion, and real-time channel estimation must be tailored to the atomic narrowband response and unique gain profile. In particular, RAQR-assisted MIMO processing, beamforming, and multi-band sensing are promising topics, as the small and non-resonant nature of vapor cells allows dense spatial deployment without mutual coupling. Further efforts are needed to linearize the RAQR’s response and mitigate quantum-induced noise through advanced coding, signal shaping, and fusion of multi-cell outputs.

\subsection{RAQR for Integrated Sensing and Communications}
\textcolor{black}{The inherent frequency agility and sensitivity of RAQRs also open the door to integrated sensing and communication (ISAC) on spaceborne platforms, yet they also pose significant self-interference challenges. A key direction is using RAQRs onboard small satellites to simultaneously support uplink reception and remote sensing based on frequency-division duplexing.} Distributed RAQR-equipped satellites may collaboratively decode weak signals from Earth terminals or measure environmental parameters without dedicated radar payloads. Their SI-traceability and polarization-resolving capability further enable precise interference analysis and electromagnetic situational awareness. A promising use case is joint radar-communication operation, where a RAQR senses echoes of communication signals for passive imaging while maintaining link performance. Designing such dual-use protocols and addressing waveform compatibility and sensitivity trade-offs will be essential.

\subsection{\textcolor{black}{Practical Challenges and Deployment Pathways}}

\textcolor{black}{Successful orbital deployment of RAQRs will require coordinated advances in hardware integration, space qualification, and mission validation. The research should address CubeSat-level SWaP constraints by advancing photonic and microfabrication technologies for miniaturized chip-scale lasers, vapor cells, and optical components. Critical engineering challenges include thermal management, radiation hardening, and structural stability under the space environment. Moreover, Doppler‑shift calibration, remote frequency tuning, and seamless insertion of RAQR front ends into standard satellite protocols need to be tackled in the future. A staged flight‑test program will begin with high‑altitude balloons and progress to orbital demonstrators to verify link budgets and end‑to‑end performance.}

\section{Conclusion}
This study demonstrated the feasibility of RAQRs as satellite front ends for 6G space-ground direct access. A review of their quantum‑electrometry principles confirmed their compatibility with space platforms. We proposed an atomic‑electronic hybrid transceiver and examined both narrowband and wideband modes. System‑level simulations reported gains in data rate, coverage, and sensing accuracy over conventional RF receivers, and a scalable architecture extended these benefits to distributed constellations. \textcolor{black}{Key challenges are outlined, including advanced signal processing, ISAC, and hardware implementation, which collectively pave the way for moving RAQR technology from lab demos to flight-ready payloads.}

\color{black}
	\section*{Acknowledgment}
     The work of Pei Xiao, Qihao Peng, and Qu Luo was supported in part by the U.K. Engineering and Physical Sciences Research Council under Grant EP/X013162/1.
    The work of Chau Yuen and Tierui Gong was supported by the Ministry of Education (MoE), Singapore, under its MoE AcRF Tier 1 Thematic Grant RT12/23 023780-00001. 
\color{black}


\bibliographystyle{IEEEtran}
\bibliography{myref}

\color{black}
\begin{IEEEbiographynophoto}{Qihao Peng} received the Ph.D. degree from Queen Mary University of London, United Kingdom, in 2024. 
        He is currently a Research Fellow in Wireless Communications in the 5GIC \& 6GIC, Institute for Communication Systems, University of Surrey, United Kingdom,. His research interests include quantum sensing for wireless communications and sensing, cell-free massive MIMO, and URLLC. 
    \end{IEEEbiographynophoto}

\begin{IEEEbiographynophoto}{Tierui Gong}
        (S'18-M'20) received the Ph.D. degree from University of Chinese Academy of Sciences (UCAS), Beijing, China, in 2020. 
        He is currently a Research Fellow with School of Electrical and Electronic Engineering, Nanyang Technological University (NTU), Singapore. His research interests include quantum information technologies for wireless communications and sensing, holographic MIMO communications, electromagnetic signal and information theory, and massive MIMO communications. 
    \end{IEEEbiographynophoto}
    
\begin{IEEEbiographynophoto}{Zihang Song} is an MSCA Postdoctoral Fellow at Connectivity (CNT) Section, Department of Electronic Systems, Aalborg University. He received the B.Sc. and M.Sc. degrees from Beihang University, China and the Ph.D. degree in Information and Communication Systems from the University of Surrey, UK. He was a research associate at King’s College London. His research focuses on developing real-time and low-power solutions for AI systems in communication networks, with a particular interest in signal processing theory, sequence models and neuromorphic computing
    \end{IEEEbiographynophoto}

\begin{IEEEbiographynophoto}{Qu Luo} is currently a Research Fellow in Wireless Communications in the 5GIC \& 6GIC, Institute for Communication Systems, University of Surrey, United Kingdom, where he received his Ph.D. degree in 2023. His research interests include proof-of-concept physical layer design, integrated sensing and communication, non-orthogonal multiple access, random access, deep/machine learning in the physical layer, and joint MAC layer and physical layer optimization. 
    \end{IEEEbiographynophoto}

\begin{IEEEbiographynophoto}{Zihuai Lin} is currently an Associate Professor at the School of Electrical and Computer Engineering at the University of Sydney, Australia. His research interests include 5G/6G cellular systems, RIS/RHS, source/channel/network coding, wireless Artificial Intelligence (AI), Artificial Intelligence of Things (AIoT) in healthcare, IoT Wireless sensing and networking, and Quantum/Ghost radar Imaging.
    \end{IEEEbiographynophoto}

\begin{IEEEbiographynophoto}{Pei Xiao} is a Professor in Wireless Communications in the Institute for Communication Systems (ICS) at University of Surrey. He is currently the technical manager of 5GIC/6GIC, leading the research team in the new physical layer work area and coordinating/supervising research activities across all the work areas.
    \end{IEEEbiographynophoto}

\begin{IEEEbiographynophoto}{Chau Yuen}
        (S'02-M'06-SM'12-F'21) received the B.Eng. and Ph.D. degrees from Nanyang Technological University, Singapore, in 2000 and 2004, respectively. Since 2023, he has been with the School of Electrical and Electronic Engineering, Nanyang Technological University. 
        He is a Distinguished Lecturer of IEEE Vehicular Technology Society, Top 2\% Scientists by Stanford University, and also a Highly Cited Researcher by Clarivate Web of Science. He has 3 US patents and published over 500 research papers at international journals or conferences.
    \end{IEEEbiographynophoto}

\color{black}

\end{document}